\documentclass[useAMS,usenatbib,usegraphicx]{mn2e}
\usepackage{amssymb,amsmath, color, url, soul, ulem}

\newcommand{\etor}{\langle B_{\rm tor}^2 \rangle}
\newcommand{\etot}{\langle B^2 \rangle}
\newcommand{\epol}{\langle B_{\rm pol}^2 \rangle}
\newcommand{\edip}{\langle B_{\rm dip}^2 \rangle}
\newcommand{\ro}{\rm Ro}
\newcommand{\msano}{{\rm M}_\odot ~{\rm yr}^{-1}}
\newcommand{\mdot}{\dot{M}}
\newcommand{\plm}{P_{\rm lm}}
\newcommand{\dplm}{\frac{{\rm d} \plm}{{\rm d} \theta}}
\newcommand{\lm}{_{\rm lm}}
\newcommand{\tp}{(\theta, \varphi)}
\newcommand{\e}[1]{\times 10^{#1}}

\title[Magnetism and the break in wind-activity relation]{Could a change in magnetic field geometry cause the  break in the wind-activity relation? }
\author[A.~A.~Vidotto et al.]{A.~A.~Vidotto$^{1}$\thanks{E-mail: Aline.Vidotto@unige.ch}, {J.-F.~Donati}$^{2,3}$, {M.~Jardine}$^{4}$, {V.~See}$^4$, {P.~Petit}$^{2,3}$,  {I.~Boisse}$^5$, \newauthor {S.~Boro Saikia}$^6$, {E.~H\'ebrard}$^{2,3}$, {S.~V.~Jeffers}$^6$, {S.~C.~Marsden}$^7$, {J.~Morin}$^8$
\\ 
$^{1}$Observatoire de l'Universit\'e de Gen\`eve, Chemin des Maillettes 51, Versoix, CH-1290, Switzerland\\
$^{2}$Universit\'e de Toulouse, UPS-OMP, IRAP, 14 avenue E. Belin, Toulouse, F-31400, France\\
$^{3}$CNRS, IRAP / UMR 5277, Toulouse, 14 avenue E. Belin, F-31400, France\\
$^{4}$SUPA, School of Physics and Astronomy, University of St Andrews, North Haugh, St Andrews, KY16 9SS, UK\\
$^{5}$LAM-UMR 7326, Aix Marseille Universit\'e, Laboratoire d'Astrophysique de Marseille, 13388, Marseille, France\\
$^{6}$Institut f\"ur Astrophysik, Georg-August-Universit\"at, Friedrich-Hund-Platz 1, D-37077, Goettingen, Germany\\
$^{7}$Computational Engineering and Science Research Centre, University of Southern Queensland, Toowoomba, 4350, Australia\\
$^{8}$LUPM-UMR5299, Universit\'e Montpellier II, Place E.~Bataillon, Montpellier, F-34095, France 
}

\date{Accepted XXX. Received YYY; in original form ZZZ}

\pubyear{2015}

\begin{document}
\label{firstpage}
\pagerange{\pageref{firstpage}--\pageref{lastpage}}
\maketitle

\begin{abstract}
Wood et al suggested that mass-loss rate is a function of X-ray flux ($\mdot \propto F_x^{1.34}$) for dwarf stars with $F_x \lesssim F_{x,6} \equiv 10^6$ erg cm$^{-2}$ s$^{-1}$. However, more active stars do not obey this relation. These authors suggested that the break at $F_{x,6}$ could be caused by significant changes in magnetic field topology that would inhibit stellar wind generation. Here, we investigate this hypothesis by analysing the stars in Wood et al's sample that had their surface magnetic fields reconstructed through Zeeman-Doppler Imaging (ZDI). Although the solar-like outliers in the $\mdot$ -- $F_x$ relation have higher fractional toroidal magnetic energy, we do not find evidence of a sharp transition in magnetic topology at $F_{x,6}$. To confirm this, further wind measurements and ZDI observations at both sides of the break are required. As active stars can jump between states with highly toroidal to highly poloidal fields, we expect significant scatter in magnetic field topology to exist for stars with $F_x \gtrsim F_{x,6}$. This strengthens the importance of multi-epoch ZDI observations. Finally, we show that there is a correlation between $F_x$ and magnetic energy, which implies that $\mdot$ -- magnetic energy relation has the same qualitative behaviour as the original $\mdot$ -- $F_x$ relation. No break is seen in any of the $F_x$ -- magnetic energy relations.
\end{abstract}

\section{Introduction}
Like the Sun, low-mass stars experience mass loss through winds during their entire lives. Although the solar wind can be probed in situ, the existence of winds on low-mass stars is known indirectly, e.g., from the observed rotational evolution of stars \citep[see][and references therein]{2014prpl.conf..433B}. Measuring the wind rates $\mdot$ of  cool, low-mass stars is not usually an easy task though. Although in their youth, when the stars are still surrounded by accretion discs, their winds have been detected \citep{2007ApJ...657..897K,2007ApJ...654L..91G}, this is no longer the case after their accretion discs dissipate. The belief is that, from the dissipation of the disc onwards, winds of low-mass stars are no longer as dense, making it more difficult to  detect them directly. 

There have been several attempts to detect radio free-free thermal emission from the winds of low-mass stars. The lack of detection means that only upper limits in $\mdot$ can be derived \citep[e.g.,][]{1997A&A...319..578V,2000GeoRL..27..501G,2014ApJ...788..112V}. Detecting X-ray emission generated when ionised wind particles exchange charges with neutral atoms of the interstellar medium has also been considered, but again only upper limits on $\mdot$ could be derived \citep{2002ApJ...578..503W}. Other methods, involving the detection of coronal radio flares \citep{1996ApJ...462L..91L} or the accretion of wind material from a cool, main-sequence component to a white dwarf component \citep{2006ApJ...652..636D,2012MNRAS.420.3281P}, have also been suggested. However, the indirect method of \citet{2001ApJ...547L..49W} has been the most successful one, enabling estimates of $\mdot$ for about a dozen dwarf stars. 

This method assumes that the observed excess absorption in the blue wing of the HI Lyman-$\alpha$ line is caused by the hydrogen wall that forms when the stellar wind interacts with the interstellar medium. Unfortunately, this method requires high-resolution UV spectroscopy, low HI-column density along the line-of-sight and a suitable viewing angle of the system \citep{2005ApJS..159..118W}, making it difficult to estimate $\mdot$ for a large number of stars. \citet{2002ApJ...574..412W,2005ApJ...628L.143W} showed that, for dwarf stars, there seems to be a relationship between the mass-loss rate per unit surface area $A_\star$ and the X-ray flux $F_x$, for stars with $F_x \lesssim 10^6$~erg~cm$^{-2}$s$^{-1}$
\begin{equation}\label{eq.wood}
\mdot/A_\star \propto F_x^{1.34}.
\end{equation}
Because stars, as they age, rotate more slowly and because their activity decreases with rotation, X-ray fluxes also decay with age \citep[e.g.,][]{2004A&ARv..12...71G}. Thus, the $\mdot$ -- $F_x$ relation implies that stars that are younger and have larger $F_x$, should also have higher $\mdot$. This, consequently, can have important effects on the early evolution of planetary systems. For example, extrapolations using Eq.~(\ref{eq.wood}) suggest that the 700~Myr-Sun would have had $\mdot$ that is $\sim 100$ times larger than the current solar mass-loss rate $\mdot_\odot = 2 \times 10^{14}~\msano$ \citep{2005ApJ...628L.143W}. If this is indeed the case, this can explain the loss of the Martian atmosphere as due to erosion caused by the stronger wind of the young Sun \citep{2004LRSP....1....2W}.

However, more active stars with $F_x\gtrsim 10^6$~erg~cm$^{-2}$s$^{-1}$ do not obey Eq.~(\ref{eq.wood}) -- their derived $\mdot$ are several orders of magnitude smaller than what Eq.~(\ref{eq.wood}) predicts \citep{2005ApJ...628L.143W,2014ApJ...781L..33W}. What could cause the change in wind characteristics at $F_x\sim 10^6$~erg~cm$^{-2}$s$^{-1}$? 
\citet{2005ApJ...628L.143W} hypothesised that the stars to the right of the ``wind dividing line'' (WDL; i.e., with $F_x\gtrsim 10^6$~erg~cm$^{-2}$s$^{-1}$) have a concentration of spots at high latitudes, as recognised in Doppler Imaging studies \citep[e.g.,][]{2009A&ARv..17..251S}, which could influence the magnetic field geometry of the star. They argue that these objects might possess a strong dipole component  (or strong toroidal fields, \citealt{2010ApJ...717.1279W}) ``that could envelope the entire star and inhibit stellar outflows''.

In this Letter, we investigate the hypothesis that solar-type stars to the right of the WDL should have a distinct magnetic field topology compared to the stars to the left of the WDL. For that, we select stars in Wood et al's sample that have large-scale surface magnetic fields previously reconstructed through Zeeman-Doppler Imaging (ZDI). In our analysis, only the large-scale surface fields are considered and we cannot assess if something happening at a smaller magnetic scale can affect the $\mdot$ -- $F_x$ relation.

\section{Magnetic field reconstruction}\label{sec.data}
Our sample consists of 7 objects (Table~\ref{table1}) out of 12 dwarf stars (not considering the Sun) in Wood et al's sample. 
%
\begin{table}
\caption{Properties of the stars considered in this work. Rotation periods $P_{\rm rot}$ were derived in the associated ZDI work (references in Table~\ref{table2}), while the remaining values are from \citealt{2004LRSP....1....2W}, \citealt{2010ApJ...717.1279W}, \citealt{2014ApJ...781L..33W}. Note that $F_x$, $\mdot$ and the magnetic maps are not necessarily derived at the same observational epoch. } \label{table1}
\begin{center}
\begin{tabular}{lccccccccccccccc}
\hline
 Star & Sp. & $R_\star$ & P$_{\rm rot}$ & $F_x$ & $\dot{M}$ &  \\ 
 ID & Type &$(R_\odot)$ & (days) & ($10^5$ erg/s/cm$^2$) & $(\dot{M}_\odot)$ & \\ \hline \hline
         EV Lac   &           M3.5V  & $      0.35$ & $       4.4$ & $       130$ & $       1.0$ & $      $ \\
    $\xi$ Boo A   &             G8V  & $      0.83$ & $       5.6$ & $        17$ & $       0.5$ & $      $ \\
    $\pi^1$ UMa   &           G1.5V  & $      0.95$ & $       5.0$ & $        17$ & $       0.5$ & $      $ \\ \hline
 $\epsilon$ Eri   &             K1V  & $      0.78$ & $      10.3$ & $       5.6$ & $      30.0$ & $      $ \\
    $\xi$ Boo B   &             K4V  & $      0.61$ & $      10.3$ & $       4.1$ & $       4.5$ & $      $ \\
       61 Cyg A   &             K5V  & $      0.68$ & $      34.2$ & $       1.0$ & $       0.5$ & $      $ \\
 $\epsilon$ Ind   &             K5V  & $      0.75$ & $      37.2$ & $      0.72$ & $       0.5$ & $      $ \\
        Sun   &             G2V  & $      1.00$ & $      25.0$ & $      0.32^a$ & $       1.0$ & $      $ \\
         \hline
\end{tabular}
\end{center}
$^a$Within its cycle, the Sun's X-ray luminosity varies from $\simeq 0.27$ to $4.7\times 10^{27}$~erg~s$^{-1}$ \citep{2000ApJ...528..537P}, corresponding to $F_x = 4.4\e4$ and $7.7\e5$~erg~cm$^{-2}$s$^{-1}$, respectively. For consistency, we chose however to use the same value as used in \citet{2014ApJ...781L..33W}, which is more representative of the Sun at minimum.
\end{table}
%
The large-scale magnetic fields were observationally-derived through the ZDI technique (\citealt{2008MNRAS.390..567M, 2012A&A...540A.138M, 2014A&A...569A..79J}, Petit et al in prep, Boro Saikia et al in prep, Boisse et al in prep). This technique consists of reconstructing the stellar surface magnetic field based on a series of circularly polarised spectra \citep{1997A&A...326.1135D}. In its most recent implementation, ZDI solves for the radial $B_r$, meridional $B_\theta$ and azimuthal $B_\varphi$ components of the stellar magnetic field, expressed in terms of spherical harmonics and their colatitude-derivatives \citep{2006MNRAS.370..629D}
\begin{equation}\label{eq.br}
B_r \tp =   \sum\lm \alpha\lm P\lm e^{im\varphi} \, ,
\end{equation}
\begin{equation}\label{eq.btheta}
B_\theta \tp =  \sum\lm  \left[ \frac{\beta\lm}{l+1} \dplm +  \frac{\gamma\lm}{l+1}\frac{im}{\sin\theta} \plm \right] e^{im\varphi} \, ,
\end{equation}
\begin{equation}\label{eq.bphi}
B_\varphi \tp = -\sum\lm \left[ \frac{\beta\lm}{l+1}\frac{im}{\sin\theta} \plm -  \frac{\gamma\lm}{l+1} \dplm \right] e^{im\varphi} \, ,
\end{equation}
where  $\alpha\lm$, $\beta\lm$, $\gamma\lm$ are the coefficients that provide the best fit to the spectropolarimetric data and $\plm \equiv \plm (\cos \theta)$ is  the associated Legendre polynomial of degree $l$ and order $m$.

To quantify the magnetic characteristics of the stars in our sample, we compute the following quantities  (Table~\ref{table2}):
\\$\bullet$ The average squared magnetic field (i.e., proportional to the magnetic energy): 
$\langle B^2 \rangle = \frac{1}{4\pi}\int (B_r^2 + B_\theta^2 + B_\varphi^2) \sin \theta d\theta d\varphi$.
\\$\bullet$ The average squared poloidal component of the magnetic field $\langle B_{\rm pol}^2 \rangle$ and its fraction $f_{\rm pol}= {\langle B_{\rm pol}^2 \rangle}/{\langle B^2 \rangle}$. To calculate $\langle B_{\rm pol}^2 \rangle$, we neglect terms with $\gamma\lm$ in Eqs.~(\ref{eq.btheta}) and (\ref{eq.bphi}).
\\$\bullet$ The axisymmetric part of the poloidal energy $\langle B_{\rm axi}^2 \rangle$ and its fraction with respect to the poloidal component $f_{\rm axi}= {\langle B_{\rm axi}^2 \rangle}/{\langle B_{\rm pol}^2 \rangle}$. To calculate $\langle B_{\rm axi}^2 \rangle$, we restrict the sum of the poloidal magnetic field energy to orders $m=0$, $m<l/2$.
\\$\bullet$  The average squared toroidal component of the magnetic field $\langle B_{\rm tor}^2 \rangle= \langle B^2 \rangle - \langle B_{\rm pol}^2 \rangle$ and its fraction $f_{\rm tor} = 1 - f_{\rm pol}$.
\\$\bullet$  The average squared dipolar component of the magnetic field $\langle B_{\rm dip}^2 \rangle$ and its fraction $f_{\rm dip}= {\langle B_{\rm dip}^2 \rangle}/{\langle B^2 \rangle}$. To calculate $\langle B_{\rm dip}^2 \rangle$, we restrict the sum of the total magnetic field energy (i.e., including the  three components of $\mathbf{B}$) to degree $l=1$. 

\begin{table*}
\caption{Magnetic properties of our sample. EV Lac, $\xi$ Boo A and $\epsilon$ Eri had their properties averaged over multi epochs (App.~A). }\label{table2}
\begin{center}
\begin{tabular}{lcccccccccl}
\hline
 Star & $\langle B^2 \rangle$ & $\langle B_{\rm pol}^2 \rangle$ & $\langle B_{\rm tor}^2 \rangle$ & $\langle B_{\rm axi}^2 \rangle$ & $\langle B_{\rm dip}^2 \rangle$ & 
$f_{\rm pol}$ & $f_{\rm tor}$ & $f_{\rm axi}$ & $f_{\rm dip}$ & Reference for surface\\ 
 ID & (G$^2$) & (G$^2$) & (G$^2$) & (G$^2$) & (G$^2$) &  &  &  &  & magnetic map \\ \hline \hline
          EV Lac$^b$  & $  3.6\e{5}$ & $  3.4\e{5}$ & $  1.7\e{4}$ & $  1.0\e{5}$ & $  2.5\e{5}$ & $ 0.95$ & $ 0.05$ & $ 0.31$ & $ 0.72 $ &            Morin et al. (2008)   \\
     $\xi$ Boo A$^b$  & $  1.8\e{3}$ & $  6.6\e{2}$ & $  1.1\e{3}$ & $  3.4\e{2}$ & $  1.2\e{3}$ & $ 0.37$ & $ 0.63$ & $ 0.51$ & $ 0.43 $ &     Morgenthaler et al. (2012)   \\
         $\pi^1$ UMa  & $  1.1\e{3}$ & $  2.0\e{2}$ & $  8.9\e{2}$ & $  3.3\e{1}$ & $  7.4\e{2}$ & $ 0.18$ & $ 0.82$ & $ 0.16$ & $ 0.68 $ &         Petit et al. (in prep)   \\ \hline
  $\epsilon$ Eri$^b$  & $  2.7\e{2}$ & $  2.0\e{2}$ & $  7.5\e{1}$ & $  7.8\e{1}$ & $  2.0\e{2}$ & $ 0.72$ & $ 0.28$ & $ 0.40$ & $ 0.75 $ &          Jeffers et al. (2014)   \\
         $\xi$ Boo B  & $  4.0\e{2}$ & $  2.7\e{2}$ & $  1.3\e{2}$ & $  7.3\e{1}$ & $  1.8\e{2}$ & $ 0.68$ & $ 0.32$ & $ 0.27$ & $ 0.45 $ &         Petit et al. (in prep)   \\
            61 Cyg A  & $  4.5\e{1}$ & $  3.9\e{1}$ & $  5.7\e{0}$ & $ 8.2\e{-1}$ & $  8.6\e{0}$ & $ 0.87$ & $ 0.13$ & $ 0.02$ & $ 0.19 $ &   Boro Saikia et al. (in prep)   \\
      $\epsilon$ Ind  & $  5.8\e{2}$ & $  5.6\e{2}$ & $  2.0\e{1}$ & $  2.8\e{2}$ & $  3.3\e{2}$ & $ 0.96$ & $ 0.04$ & $ 0.51$ & $ 0.56 $ &        Boisse et al. (in prep)   \\
                \hline
\end{tabular}
\end{center}
\end{table*}

Similarly to the Sun, it has been recognised that stellar magnetism can evolve on a yearly timescale, with some stars exhibiting complete magnetic cycles. Three of the objects in our sample, namely EV Lac, $\xi$ Boo A and $\epsilon$ Eri, have multi-epoch reconstructed surface maps. For these stars, $f_{\rm pol}$, $f_{\rm axi}$, $f_{\rm dip}$ and $\langle B^2 \rangle$, which are shown in Table~\ref{table2}, were averaged over multiple observing epochs (see Appendix A). We note that, as the observations are not done at regular time intervals, the time the star spends in a given `magnetic state' might not be well represented by this simple average. However, we believe that this approach is a better representation of the magnetic characteristic for each of these stars over the choice of one single-epoch map, in particular when the star-to-star differences are comparable to the year-to-year (i.e., amplitude) variations that stars exhibit (more important in the most active stars).

\section{Magnetic fields as the cause of the break in the $\mdot$ -- $F_x$ relation}\label{sec.results}
One proposed idea for the break in the $\mdot$ -- $F_x$ relation for the most active stars is that stars that are to the right of the WDL have surface magnetic field topologies that are significantly different, with either stronger dipolar \citep{2005ApJ...628L.143W} or toroidal fields \citep{2010ApJ...717.1279W}, that partially inhibit the outflow of the stellar wind, giving rise to reduced $\mdot$. To verify the first hypothesis, we compute $\langle B_{\rm dip}^2 \rangle$ and $f_{\rm dip}$ for all the stars in our sample (Table~\ref{table2}). We do not find any particular evidence that stars to the right of the WDL, namely $\pi^1$ UMa, $\xi$ Boo A and EV Lac, have dipolar magnetic fields whose characteristics are remarkably different from the remaining stars in our sample. 

\begin{figure*}
	\includegraphics[width=160mm]{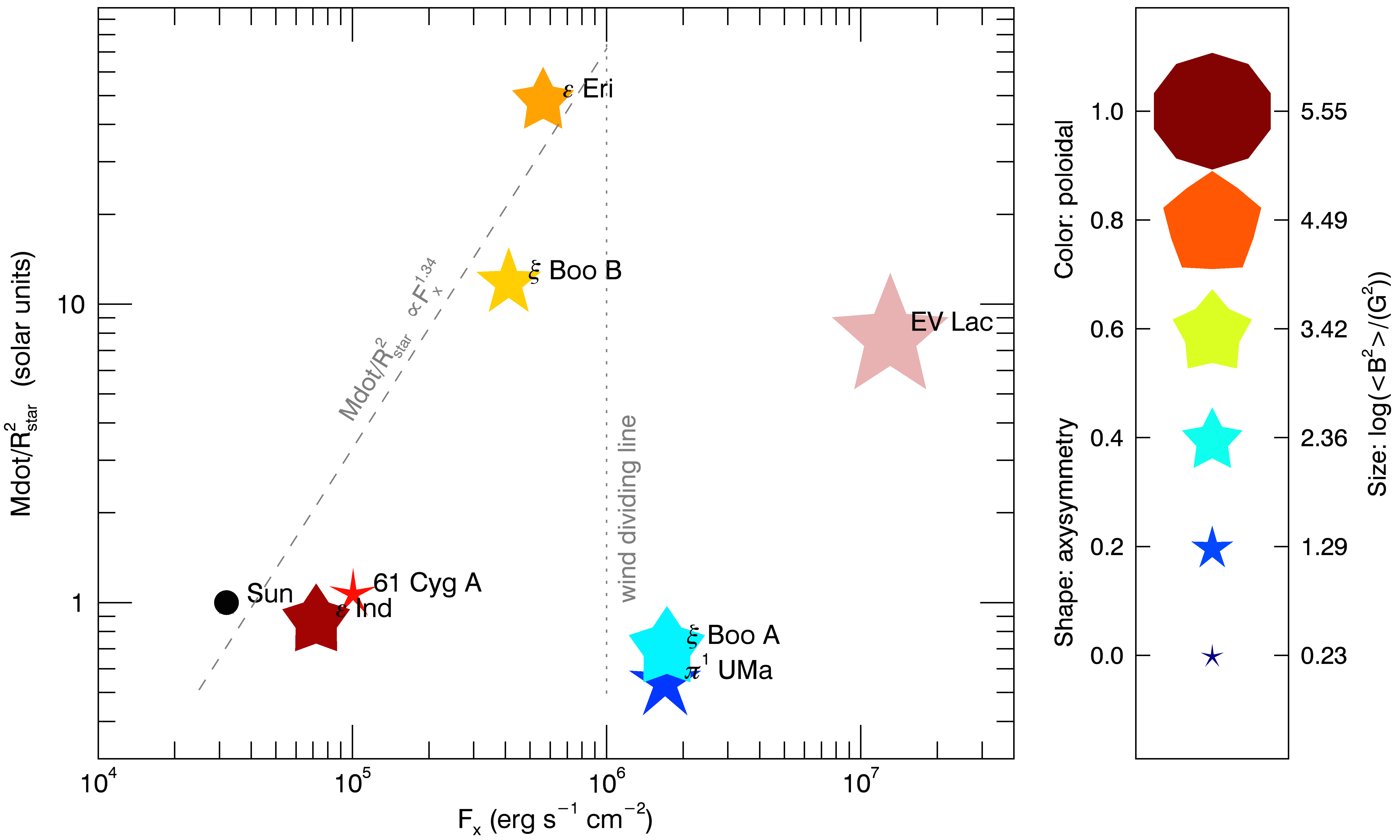}
\caption{Mass-loss rates per surface area versus X-ray fluxes showing only the stars for which we have reconstructed surface fields. Symbol sizes are proportional to $\log \langle B^2 \rangle$, their colours indicate the fractional poloidal energy (ranging from deep red for purely poloidal field $f_{\rm pol}=1$ to blue for purely toroidal field $f_{\rm pol}=0$), and their shapes indicate the fraction of axisymmetry of the poloidal component (ranging from a decagon for purely axisymmetric field $f_{\rm axi}=1$ to a point-shaped star for $f_{\rm axi}=0$). The $\mdot$ -- $F_x$ relation (Eq.~\ref{eq.wood}) is shown as a dashed line and the WDL at $F_x=10^6$~erg~cm$^{-2}$s$^{-1}$ is shown as a dotted line. The Sun (filled black circle) is shown for comparison. }\label{fig.wind_confusogram}
\end{figure*}

Fig.~\ref{fig.wind_confusogram} presents a similar version of Wood et al's diagram ($\mdot/A_\star$ versus $F_x$) showing only the stars for which we have reconstructed surface fields. The symbols are as in Fig.\,3 of \citet{2009ARA&A..47..333D}, in which their sizes are proportional to $\log \langle B^2 \rangle$, their colours are related to $f_{\rm pol}$, and their shapes to $f_{\rm axi}$ (see caption of figure). EV Lac is included in this plot for completeness but it is not considered in the analysis that follows. It has been argued that EV Lac could be a discrepant data point in the wind-activity relation because it is the least solar-like star in the sample \citep{2004LRSP....1....2W}. Indeed, it has been revealed that the magnetism of active M dwarf stars, like EV Lac, has striking differences from solar-like objects, both in intensity and geometry \citep{2006Sci...311..633D,2008MNRAS.390..545D,2007ApJ...656.1121R,2009ApJ...692..538R,2008MNRAS.390..567M,2010MNRAS.407.2269M}. Note that, except for EV Lac, our objects have Rossby numbers $\ro > 0.1$, where $\ro$ is defined as the ratio between the convective turnover timescale to the rotation period of the star ($\ro$  compiled by \citealt{2014MNRAS.441.2361V}). Therefore, none of our solar-type stars are in the X-ray saturated regime \citep{2003A&A...397..147P}.

\begin{table}
\caption{Slope of the ordinary least square fits to the logarithm of our data (i.e., $y \propto x^{\rm slope}$).
\label{tab.slopes}}
\begin{center}
\begin{tabular}{lccccc}
\hline
y(x) & $F_x (\etor)$ & $F_x (\epol)$ & $F_x (\edip)$ & $F_x (\etot)$ \\
slope & $0.65 \pm 0.06$& $0.47 \pm 0.05$ & $0.50 \pm 0.03$ & $0.55 \pm 0.04$ \\
\hline
\end{tabular}
\end{center}
\end{table}

Fig.~\ref{fig.wind_confusogram} shows no evidence that $\mdot$ and $f_{\rm axi}$ (symbol shape) are related, indicating that the axisymmetry of the poloidal field cannot explain the  wind-activity relation nor its break. In the Sun, $\mdot$ and $f_{\rm axi}$ also seem to be unrelated. The solar dipolar field, nearly aligned with the Sun's rotation axis (and therefore nearly axisymmetric) at minimum activity phase, increasingly tilts towards the equator as the cycle approaches activity maximum \citep{2012ApJ...757...96D}, decreasing therefore $f_{\rm axi}$. In spite of the cycle variation of $f_{\rm axi}$ in the Sun, the solar wind mass-loss rates remains largely unchanged \citep{2010ApJ...715L.121W}. 
 
What stands out in Fig.\,\ref{fig.wind_confusogram} is that solar-type stars with larger toroidal fractional fields (bluer) are concentrated to the right of the WDL. With the limited number of stars with both ZDI and wind measurements, there is no evidence at present for a sharp transition in magnetic topology across the WDL. Instead, the concentration of stars with dominantly toroidal fields to the right of the WDL reflects a trend towards more strongly toroidal fields with increasing $F_x$, which is in turn associated with rapid rotation. \citet{2008MNRAS.388...80P} showed that, as stellar rotation increases, solar-type stars tend to have higher fraction of toroidal fields (symbols becoming increasingly bluer in diagrams such as the one expressed in Fig.~\ref{fig.wind_confusogram}; see also Fig.~3 in \citealt{2009ARA&A..47..333D}). Since rotation is linked to activity, it is not surprising that $\pi^1$ UMa and $\xi$ Boo A, the most active and rapidly rotating solar-type stars in our sample, are the ones with the largest fraction of toroidal fields. To confirm our findings, it is essential to perform further wind measurements and ZDI observations of more stars at both sides of the break.

\begin{figure}
	\centering
	\includegraphics[width=80mm]{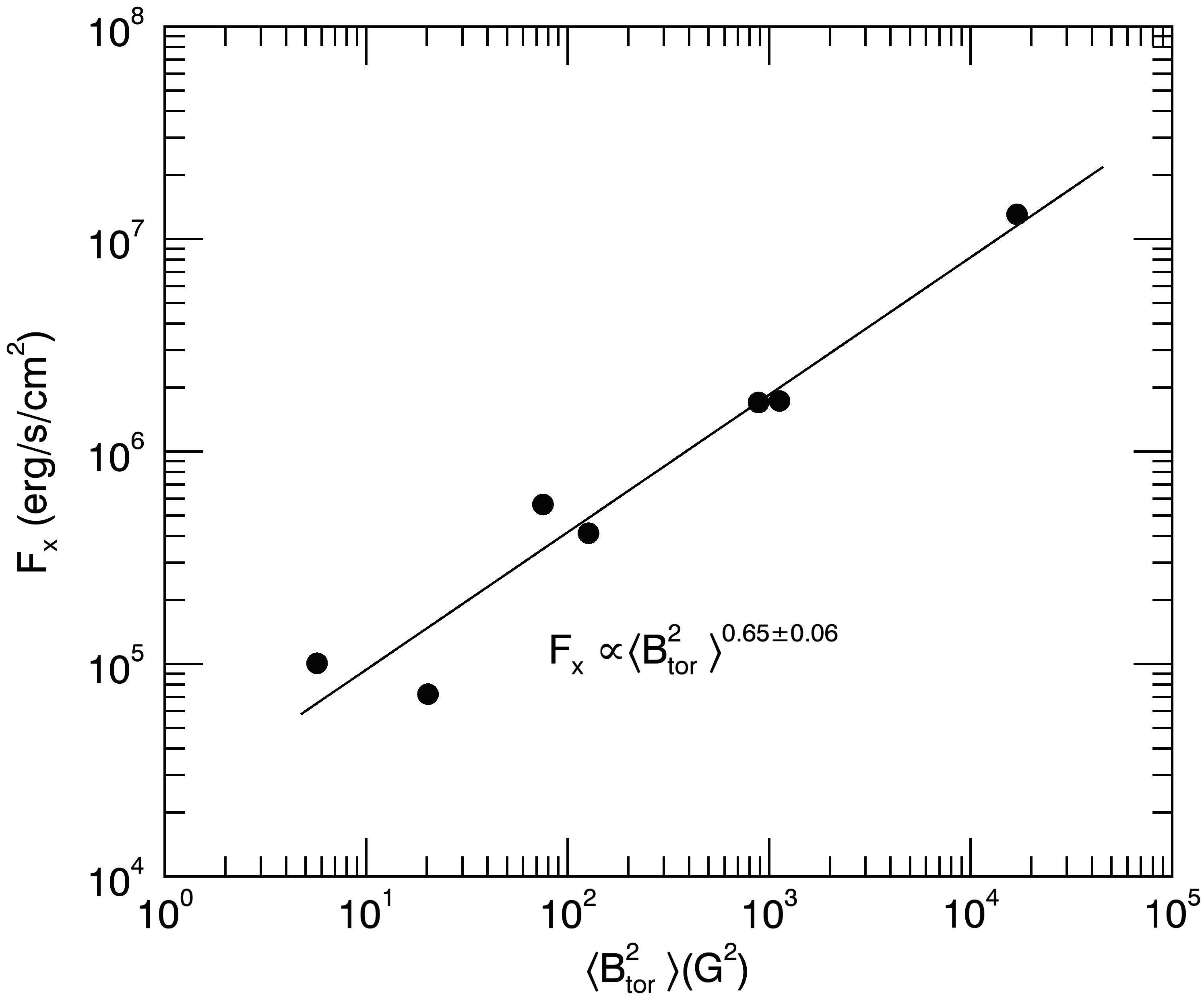}
\caption{Relation between $F_x$ and $\etor$ does not show a break at $F_x \sim 10^6$~erg~cm$^{-2}$s$^{-1}$. Solid line is a power-law fit the data. Because $\etor$ and $\epol$ are correlated \citep{see2015}, correlations between $F_x$ and $\epol$ or $\etot$ also exist (see Table~\ref{tab.slopes}). }\label{fig.Fx_Etor}
\end{figure}

We note also that if instead of plotting $F_x$ in the $x$-axis of Fig.~\ref{fig.wind_confusogram}, we plotted $\etor$, this plot would present the same general property: $\mdot/A_\star$ that increases with  $\etor$ for the least active stars and then a break in this relation for $\pi^1$ UMa and $\xi$ Boo A, with the break  occurring somewhere between $\etor \sim 200$ and  $800$~G$^2$.
{\footnote{{The same characteristic is seen in a plot of $\mdot/A_\star$ vs $1/\ro$, since $F_x$ and $\ro$ are anti-correlated \citep[e.g.,][]{1994A&A...292..191S}}.} }
The similar characteristics between $\mdot/A_\star$ vs $F_x$ and $\mdot/A_\star$ vs $\etor$  is because $\etor$ is correlated to $F_x$, as shown in Fig.~\ref{fig.Fx_Etor} (this figure includes EV Lac). Because $\etor$ and $\epol$ are correlated \citep{see2015}, correlations between $F_x$ and $\epol$ or $\etot$ also exist (see Table~\ref{tab.slopes}). No break is seen in any of the $F_x$ -- magnetic energy relations. 
Because magnetism and X-ray flux evolve on $\sim$yearly timescales, intrinsic variability is expected to increase the spread in all the relations in Table~\ref{tab.slopes}, further reinforcing the importance of multi-epoch ZDI observations.

Another important point to consider is that very active stars can jump between states with highly toroidal fields and mostly poloidal fields (\citealt{2009A&A...508L...9P, 2012A&A...540A.138M, 2015A&A...573A..17B}). Therefore, if the high fraction of toroidal magnetic fields is related to the break on the $\mdot$ -- $F_x$ relation, there may be some significant scatter in magnetic field topology on the right of the WDL. E.g.,  during $\sim$ 5 yr of observations, the fractional toroidal field of $\xi$ Boo A varied between $f_{\rm tor } \simeq 33\%$ to $81\%$, with an average of $63\%$, giving this star a blueish color in Fig.~\ref{fig.wind_confusogram} (Appendix A). If this star were observed only during an epoch where its field is dominantly poloidal, the trend that stars to the right of the WDL are more toroidal (bluer) would not be recovered. This further strengthens the need of multi-epoch ZDI observations (cf. Section \ref{sec.data}).  

Assuming that a change in magnetic field topology is such that it will significantly affect the mass-loss rates of stellar winds, it would still take on the order of a year (depending on the size of the astrosphere) for stellar winds to propagate out into the astrosphere boundary, where the Ly-$\alpha$ photons are absorbed. Therefore, simultaneous derivations of mass-loss rate (through Ly-$\alpha$ absorption) and magnetic field topology (through ZDI) are likely not going to be linked to each other. It would, nevertheless, be interesting to monitor both Ly-$\alpha$ absorption and  magnetic topologies to see whether variations of the latter mimic those of the former with a $\sim 1$~yr delay.

\section{Conclusions and Discussion}\label{sec.conclusion}
In this letter, we have investigated if stellar magnetic fields could be the cause of the break in the wind-activity ($\mdot$ -- $F_x$) relation, for stars with $F_x\gtrsim 10^6$~erg~cm$^{-2}$s$^{-1}$ \citep{2005ApJ...628L.143W}. Our sample consisted of stars observed by Wood et al that also had observationally-derived large-scale magnetic fields (\citealt{2008MNRAS.390..567M, 2012A&A...540A.138M, 2014A&A...569A..79J}, Petit et al in prep, Boro Saikia et al in prep, Boisse et al in prep). 

\citet{2005ApJ...628L.143W} and \citet{2010ApJ...717.1279W} suggested that the break in the $\mdot$ -- $F_x$ relation seen for the most active stars could be caused by the presence of strong dipolar or toroidal fields that would inhibit the wind generation. 
In our analysis, we did not find any particular evidence that the dipolar field characteristics (or the degree of axisymmetry of the poloidal field) change at the WDL. We found that solar-type stars to the right of the WDL (namely $\xi$ Boo A and $\pi^1$ UMa) have higher fractional toroidal fields (blueish points in Fig.~\ref{fig.wind_confusogram}), but no break or sharp transition is found (Fig.~\ref{fig.Fx_Etor}, \citealt{2008MNRAS.388...80P}).

We also showed that there is a correlation between $F_x$ and magnetic energy (Table~\ref{tab.slopes}), which implies that $\mdot$ -- magnetic energy relation has the same general properties as the original $\mdot$ -- $F_x$ relation (i.e., a positive correlation for $F_x \lesssim 10^6$~erg~cm$^{-2}$s$^{-1}$, followed by a break). Contrary to the $\mdot$ -- $F_x$ relation, no break at the WDL  is seen in the $F_x$ -- magnetic energy relations. 

We note that very active stars can jump between states with highly toroidal fields and mostly poloidal fields (\citealt{2009A&A...508L...9P, 2012A&A...540A.138M, 2015A&A...573A..17B}). If magnetic fields are actually affecting stellar winds (i.e., the high fraction of toroidal fields is related to the break in the $\mdot$ -- $F_x$ relation), there may be some significant scatter in magnetic field topology on the right of the WDL, even for a given star observed at various phases of its magnetic cycle.  

We finally add that the break in the $\mdot$ -- $F_x$ relation found by \citet{2005ApJ...628L.143W} is at odds with some theoretical works on stellar winds and empirically-derived relations used to compute outflow rates ($\mdot_{\rm CME}$) of coronal mass ejections (CMEs). The time-dependent stellar wind models of \citet{2007A&A...463...11H, 2014A&A...570A..99S,2015A&A...577A..28J} predict a different (in general, weaker) dependence of $\mdot$ with $F_x$ than that of Eq.~(\ref{eq.wood}) and do not predict a sharp decrease in $\mdot$ for stars to the right of the WDL. These models use empirical data on rotation, wind temperature, density and magnetic field strength to constrain theoretical wind scenarios. They are, however, limited to one or two dimensions and cannot, therefore, incorporate the 3D nature of stellar magnetic fields. 3D stellar wind studies that incorporates observed stellar magnetic maps have been carried out but they are currently not time-dependent \citep{2010ApJ...719..299C, 2012MNRAS.423.3285V,2014MNRAS.438.1162V, 2013MNRAS.431..528J}.

Other theoretical works suggested that CMEs could be the dominant form of mass loss for active stars \citep{2012ApJ...760....9A,2013ApJ...764..170D, 2015arXiv150604994O}. These works predicted that $\mdot_{\rm CME} \gtrsim 100 \mdot_\odot$ for active solar-like stars, in contradiction to the results found by Wood et al for active stars such as $\pi^1$ UMa and $\xi$ Boo A. A possibility for this disagreement is that the solar CME-flare relation cannot be extrapolated all the way to active stars \citep{2013ApJ...764..170D}.  \citet{2013ApJ...764..170D}  suggested that there might not be a one-to-one relation between observed stellar X-ray flares and CMEs. They argue that on active stars, CMEs are more strongly confined, and so fewer of them are produced for any given number of X-ray flares. This stronger confinement could be caused by the noticeable differences between the solar and stellar magnetic field characteristics. What we have shown here is that indeed the most active stars seem to have more toroidal large-scale magnetic field topologies. Numerical modelling efforts would hopefully be able to shed light on whether the large-scale toroidal fields could indeed result in more confined CMEs. 

\section*{Acknowledgements}
AAV acknowledges support from the Swiss National Science Foundation through an Ambizione Fellowship. SVJ and SBS  acknowledge research funding by the DFG under grant SFB 963/1, project A16. The authors thank Dr Rim Fares for useful discussions and the anonymous referee for constructive comments and suggestions.

\def\aj{{AJ}}                   
\def\araa{{ARA\&A}}             
\def\apj{{ApJ}}                 
\def\apjl{{ApJ}}                
\def\apjs{{ApJS}}               
\def\ao{{Appl.~Opt.}}           
\def\apss{{Ap\&SS}}             
\def\aap{{A\&A}}                
\def\aapr{{A\&A~Rev.}}          
\def\aaps{{A\&AS}}              
\def\azh{{AZh}}                 
\def\baas{{BAAS}}               
\def\jrasc{{JRASC}}             
\def\memras{{MmRAS}}            
\def\mnras{{MNRAS}}             
\def\pra{{Phys.~Rev.~A}}        
\def\prb{{Phys.~Rev.~B}}        
\def\prc{{Phys.~Rev.~C}}        
\def\prd{{Phys.~Rev.~D}}        
\def\pre{{Phys.~Rev.~E}}        
\def\prl{{Phys.~Rev.~Lett.}}    
\def\pasp{{PASP}}               
\def\pasj{{PASJ}}               
\def\qjras{{QJRAS}}             
\def\skytel{{S\&T}}             
\def\solphys{{Sol.~Phys.}}      
\def\sovast{{Soviet~Ast.}}      
\def\ssr{{Space~Sci.~Rev.}}     
\def\zap{{ZAp}}                 
\def\nat{{Nature}}              
\def\iaucirc{{IAU~Circ.}}       
\def\aplett{{Astrophys.~Lett.}} 
\def\apspr{{Astrophys.~Space~Phys.~Res.}}   
\def\bain{{Bull.~Astron.~Inst.~Netherlands}}    
\def\fcp{{Fund.~Cosmic~Phys.}}  
\def\gca{{Geochim.~Cosmochim.~Acta}}        
\def\grl{{Geophys.~Res.~Lett.}} 
\def\jcp{{J.~Chem.~Phys.}}      
\def\jgr{{J.~Geophys.~Res.}}    
\def\jqsrt{{J.~Quant.~Spec.~Radiat.~Transf.}}   
\def\memsai{{Mem.~Soc.~Astron.~Italiana}}   
\def\nphysa{{Nucl.~Phys.~A}}    
\def\physrep{{Phys.~Rep.}}      
\def\physscr{{Phys.~Scr}}       
\def\planss{{Planet.~Space~Sci.}}           
\def\procspie{{Proc.~SPIE}}     
\def\actaa{{Acta~Astronomica}}     
\def\pasa{{Publications of the ASA}}     
\def\na{{New Astronomy}}     

\let\astap=\aap
\let\apjlett=\apjl
\let\apjsupp=\apjs
\let\applopt=\ao
\let\mnrasl=\mnras



\appendix
\section{Average magnetic characteristics from multi-epoch studies (online only)}\label{sec.apA}
All the relations involving activity and magnetism have as common characteristics a relatively high dispersion. This dispersion is likely a result of the evolution of stellar magnetism, which happens in a range of different timescales, from short ones, on the order of days to months, due to the evolution of spots, to medium timescales, on the order of years to decades, due to cycles, and finally on stellar evolutionary timescales. In this work, whenever possible, we used reconstructed stellar surface magnetic maps of multi-epoch observing campaigns. Table \ref{table_ap} lists the characteristics of the reconstructed magnetic field at individual epochs for  $\xi$ Boo A, $\epsilon$ Eri and EV Lac. The values of $f_{\rm pol}$, $f_{\rm axi}$, $f_{\rm dip}$ and $\langle B^2 \rangle$ presented in Table 2 were derived by averaging the respective quantities presented in Table \ref{table_ap}: i.e., values in Table 2 for these these three stars are in fact $\overline{f_{\rm pol}}$, $\overline{f_{\rm axi}}$, $\overline{f_{\rm dip}}$ and $\overline{\langle B^2 \rangle}$, respectively. The remaining quantities in Table 2 are then calculated as follows:  $\epol =\overline{f_{\rm pol}} \, \overline{\langle B^2 \rangle}$, $\etor =(1-\overline{f_{\rm pol}}) \, \overline{\langle B^2 \rangle}$, $\langle B_{\rm dip}^2 \rangle =\overline{f_{\rm dip}} \, \overline{\langle B^2 \rangle}$ and $\langle B_{\rm axi}^2 \rangle = \overline{f_{\rm axi}} \, \overline{f_{\rm pol}} \, \overline{\langle B^2 \rangle}$.

\begin{table*}
\caption{Magnetic properties of the stars in our sample that have multi-epoch magnetic field reconstructions. The results for $\xi$ Boo A were presented in \citet{2012A&A...540A.138M}, for $\epsilon$ Eri  in \citet{2014A&A...569A..79J} and for EV Lac in  \citet{2008MNRAS.390..567M}.} \label{table_ap}
\begin{center}
\begin{tabular}{lccccccccccccccc}
\hline
Epoch & $\langle B^2 \rangle $ & $\langle B_{\rm pol}^2 \rangle $ & $\langle B_{\rm tor}^2 \rangle $ & $\langle B_{\rm axi}^2 \rangle $ & $\langle B_{\rm dip}^2 \rangle $ & $f_{\rm pol}$ & $f_{\rm tor}$ & $f_{\rm axi}$ & $f_{\rm dip}$\\ 
& (G$^2$)& (G$^2$)& (G$^2$)& (G$^2$)& (G$^2$)&&&&&\\
\hline \hline
{\bf $\xi$ Boo A:} \\ 
2007/08  & $  4.7\e{3}$ & $  8.8\e{2}$ & $  3.8\e{3}$ & $  1.4\e{2}$ & $  3.9\e{3}$ & $  0.19$ & $  0.81$ & $  0.16$ & $  0.82 $ \\
2008/02  & $  7.5\e{2}$ & $  4.6\e{2}$ & $  3.0\e{2}$ & $  2.8\e{2}$ & $  3.2\e{2}$ & $  0.60$ & $  0.40$ & $  0.60$ & $  0.43 $ \\
2009/06  & $  1.8\e{3}$ & $  7.1\e{2}$ & $  1.1\e{3}$ & $  3.5\e{2}$ & $  1.1\e{3}$ & $  0.39$ & $  0.61$ & $  0.49$ & $  0.61 $ \\
2010/01  & $  1.1\e{3}$ & $  3.8\e{2}$ & $  7.2\e{2}$ & $  2.3\e{2}$ & $  6.8\e{2}$ & $  0.35$ & $  0.65$ & $  0.60$ & $  0.62 $ \\
2010/06  & $  8.5\e{2}$ & $  5.7\e{2}$ & $  2.8\e{2}$ & $  3.4\e{2}$ & $  4.2\e{2}$ & $  0.67$ & $  0.33$ & $  0.58$ & $  0.49 $ \\
2010/08  & $  1.6\e{3}$ & $  3.0\e{2}$ & $  1.3\e{3}$ & $  2.6\e{2}$ & $  1.2\e{3}$ & $  0.19$ & $  0.81$ & $  0.88$ & $  0.74 $ \\
2011/01  & $  1.7\e{3}$ & $  3.5\e{2}$ & $  1.4\e{3}$ & $  9.2\e{1}$ & $  1.4\e{3}$ & $  0.20$ & $  0.80$ & $  0.27$ & $  0.82 $ \\
average: & $  1.8\e{3}$ & &  &  &  & $ 0.37$ & $ 0.63$ & $ 0.51$ & $ 0.43 $  \\
\hline 
{\bf $\epsilon$ Eri:} \\ 
2007/01  & $  1.9\e{2}$ & $  1.7\e{2}$ & $  1.5\e{1}$ & $  1.7\e{1}$ & $  1.7\e{2}$ & $  0.92$ & $  0.08$ & $  0.10$ & $  0.90 $ \\
2008/01  & $  1.2\e{2}$ & $  1.1\e{2}$ & $  7.7\e{0}$ & $  6.8\e{1}$ & $  7.4\e{1}$ & $  0.94$ & $  0.06$ & $  0.60$ & $  0.62 $ \\
2010/01  & $  3.2\e{2}$ & $  1.3\e{2}$ & $  1.9\e{2}$ & $  4.8\e{1}$ & $  2.1\e{2}$ & $  0.41$ & $  0.59$ & $  0.38$ & $  0.67 $ \\
2011/10  & $  1.2\e{2}$ & $  9.2\e{1}$ & $  3.2\e{1}$ & $  5.7\e{1}$ & $  8.3\e{1}$ & $  0.74$ & $  0.26$ & $  0.62$ & $  0.67 $ \\
2012/10 & $  4.3\e{2}$ & $  2.4\e{2}$ & $  1.9\e{2}$ & $  1.1\e{2}$ & $  3.2\e{2}$ & $  0.55$ & $  0.45$ & $  0.46$ & $  0.75 $ \\
2013/10  & $  4.6\e{2}$ & $  3.6\e{2}$ & $  9.9\e{1}$ & $  7.9\e{1}$ & $  3.8\e{2}$ & $  0.78$ & $  0.22$ & $  0.22$ & $  0.84 $ \\
average:& $  2.7\e{2}$ & &  && & $ 0.72$ & $ 0.28$ & $ 0.40$ & $ 0.75 $ \\
\hline
{\bf EV Lac:} \\
2006/08  & $  4.1\e{5}$ & $  3.8\e{5}$ & $  3.2\e{4}$ & $  1.1\e{5}$ & $  2.8\e{5}$ & $  0.92$ & $  0.08$ & $  0.30$ & $  0.68 $ \\
2007/08  & $  3.1\e{5}$ & $  3.0\e{5}$ & $  5.2\e{3}$ & $  9.4\e{4}$ & $  2.2\e{5}$ & $  0.98$ & $  0.02$ & $  0.31$ & $  0.72 $ \\
average: & $  3.6\e{5}$ &  &  &  &  & $ 0.95$ & $ 0.05$ & $ 0.31$ & $ 0.72 $ \\
\hline
\end{tabular}
\end{center}
\end{table*}

%
\begin{figure*}
	\includegraphics[width=150mm]{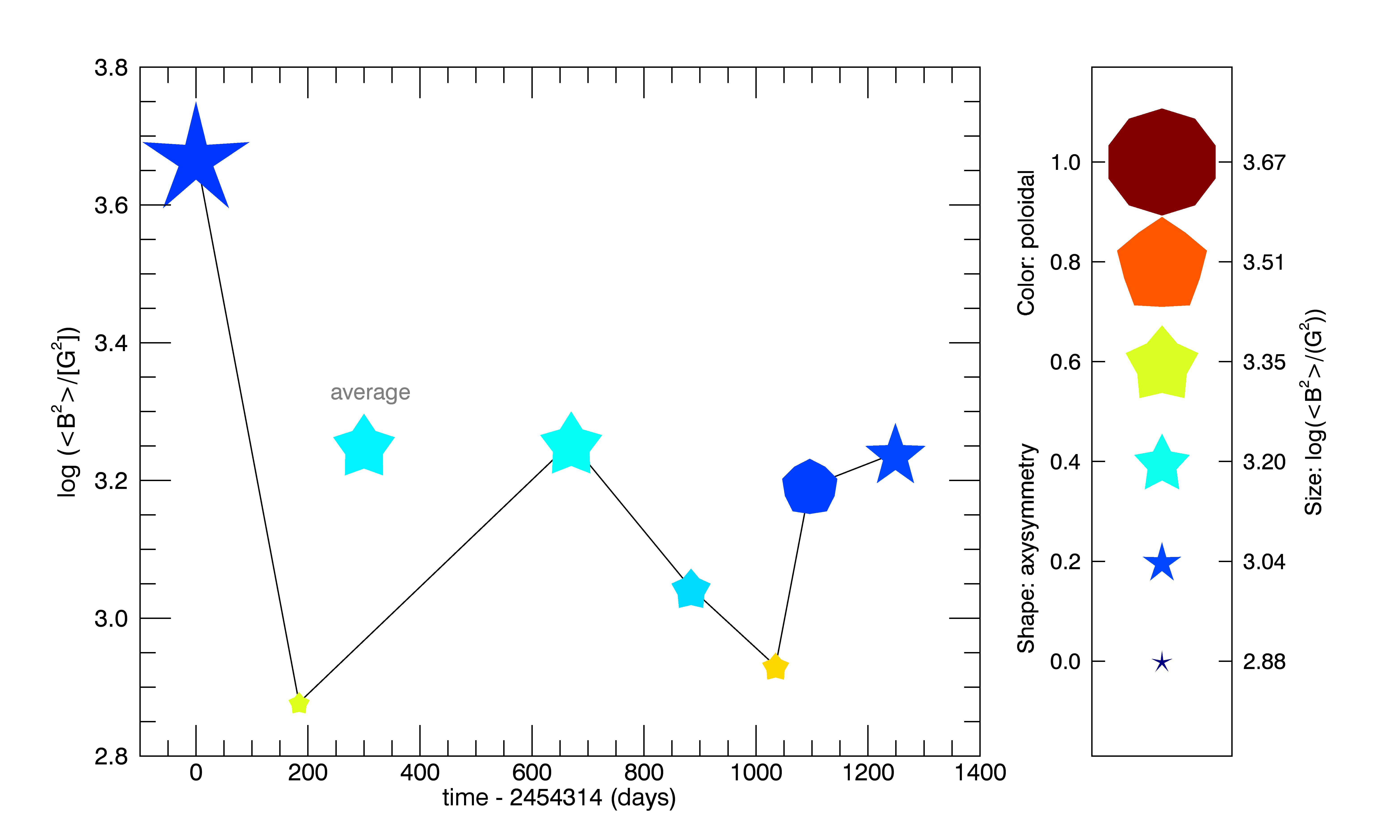}
	\caption{Evolution of the magnetic field properties of $\xi$~Boo~A, according to multi-epoch magnetic field reconstructions from \citet{2012A&A...540A.138M}. A better representation of the magnetic field property is calculated as average over the observed epochs (see text). We believe this approach is a better representation of the magnetic characteristic of the star over the choice of one single-epoch map. The same approach is employed for all the stars in our sample with multi-epoch observations, namely EV Lac and $\epsilon$ Eri. The solid line is a connector and does not represent any particular fit to the data.}\label{fig.average_confusogram}
\end{figure*}

\bsp
\label{lastpage}
\end{document}